# Strain Effect on Air-Stability of Monolayer $CrSe_2$

Jun Chen[1, †], Linwei Zhou[2, †], Nanshu Liu[1], Jingsi Qiao[3], Xieyu Zhou[1], Cong Wang[1, *], and Wei Ji[1, *]

[1]*Beijing Key Laboratory of Optoelectronic Functional Materials & Micro-Nano Devices, Department of Physics, Renmin University of China, Beijing 100872, P.R. China*

[2]*College of Physics and Optoelectronic Engineering, Shenzhen University, Shenzhen, 518060, P.R. China*

[3]*School of Information and Electronics, MIIT Key Laboratory for Low-Dimensional Quantum Structure and Devices, Beijing Institute of Technology, Beijing 100081, P.R. China*

\* Corresponding authors: C.W. (email: wcphys@ruc.edu.cn), W.J. (email: wji@ruc.edu.cn)

† These authors contributed equally to this work.

## Abstract

The discovery of two dimensional (2D) magnetic materials has brought great research value for spintronics and data storage devices. However, their air-stability as well as the oxidation mechanism has not been unveiled, which limits their further applications. Here, by first-principles calculations, we carried out a detailed study on the oxidation process of monolayer $CrSe_2$ and biaxial tensile strain effect. We found dissociation process of $O_2$ on pristine $CrSe_2$ sheet is an endothermic reaction with a reaction energy barrier of 0.53 eV, indicating its thermodynamics stability. However, such a process becomes exothermic under a biaxial tensile strain reaching 1%, accompanying with a decreased reaction barrier, leading to reduced stability. These results manifest that in-plane strain plays a significant role in modifying air-stability in $CrSe_2$ and shed considerable light on searching appropriate substrate to stabilize 2D magnetic materials.

## Introduction

2D magnetic materials have attracted considerable interest due to their wide application value in sensors[1], magnetic storage[2], and other technologies[3-7]. At present, intrinsic ferromagnetism have been observed in 2D materials, such as $CrI_3$[8-10], $Cr_2Ge_2Te_6$[11, 12], $Fe_3GeTe_2$[13, 14]. Their magnetic properties can be tuned by layer number[8, 11], charge doping[15], strain[16], and stacking orders[17], which holds promise for control of magnetism in functional devices. However, most 2D magnetic materials do not show persistent air and water stability up to now and have to be encapsulated. For example, $CrI_3$ nanosheet degrades in air in 15 minutes[18] and ferromagnetic signals of $Fe_3GeTe_6$ nanosheet vanish under the atmosphere for a few hours[13], which limit their further application. Physical encapsulation offers advantages of direct device integration but lacks high scalability and possibility of additional chemical functionalization[19]. Thus, air-stability is critical for both fundamental studies and technological applications of 2D magnetic materials[8, 11, 13].

Compared with previously discovered 2D magnetic materials, $CrSe_2$ nanosheet can maintain good air-stability for several months[20]. $FePS_3$[21], $CrSBr$[22], and $CrTe_2$[23] also exhibit good stability in air, but mechanism of their stability is yet to be unveiled. Therefore, exploring the origin of high air-stability of 2D magnetic materials and founding possible ways improving their stability become a key problem to be solved. Epitaxial strain induced by substrate is common during materials fabrication process and could significantly manipulates ether intralayer and interlayer magnetism of $CrSe_2$ and $CrTe_2$[16, 20, 23]. Whether the common existing epitaxy strain is capable of tuning air-stability of 2D magnets is yet to be unveiled.

To explore the mechanism of air-stability in monolayer $CrSe_2$ and tensile strain effect on stability, we studied $O_2$ dissociation process on $CrSe_2$ using first-principles calculations. For pristine $CrSe_2$, the dissociative adsorption of $O_2$ is endothermic with an energy gaining of 0.08 eV. The energetically unfavored surface oxidation reaction could be one of the reasons for the exceptional air-stability of monolayer $CrSe_2$. Meanwhile, we found that biaxial tensile strain reduces the air-stability of $CrSe_2$. Upon biaxial tensile strain, the dissociation reaction of $O_2$ becomes an exothermic reaction and energy barrier reduces from 0.52 eV to 0.20 eV. Due to a largely stabilized final state, the dissociation process changes from endothermic to exothermic transition under a strain larger than 1%. Biaxial tensile strain in the range from 0 to 4% induces increase of the charge transfer from 0.42 to 0.52 e, which could activate the O-O bond and facilitate its rupture, leading to the reduction of energy barriers.

## Methods

All calculations were performed on the basis of the density functional theory (DFT) framework. The projector augmented wave (PAW) method and the Perdew-Burke-Ernzerhof (PBE) functional under generalized gradient approximation (GGA) were adopted[24-26], as implemented in the Vienna *ab initio* simulation package (VASP) code[27]. The DFT-D3 method of Grimme was used to evaluate the contributions from the van der Waals interactions[28]. The geometry optimization was performed until the Hellmann-Feynman force acting on per atom was less than 0.02 eV/Å, the energy convergence criterion was set to $1\times10^{-4}$ eV. The kinetic energy cut-off for plane-wave basis set was set to 400 eV. The on-site Coulomb interaction to the Cr d orbitals had U and J values of 4.6 eV and 0.6 eV, respectively, as revealed by a linear response method[29, 30] and comparison with the results of HSE06 functional[31]. These values are comparable to those adopted in modelling $CrI_3$[32], $CrS_2$[15], and $CrSe_2$[30]. A vacuum region of 20 Å was added to the perpendicular direction to eliminate interactions between periodic images. A $k$-mesh of 4×4×1 was employed in geometric optimization. The climbing-image nudged elastic band (CI-NEB) method[33, 34] was used to determine the adsorption and dissociation path of $O_2$ molecule. A $k$-mesh of 2×2×1 was employed in NEB calculations. The intermediate images of each CI-NEB simulation were relaxed until the perpendicular forces are smaller than 0.05 eV/Å. The number of electrons transferred from monolayer $CrSe_2$ to $O_2$ was calculated by Bader analysis[35].

## Results and Discussions

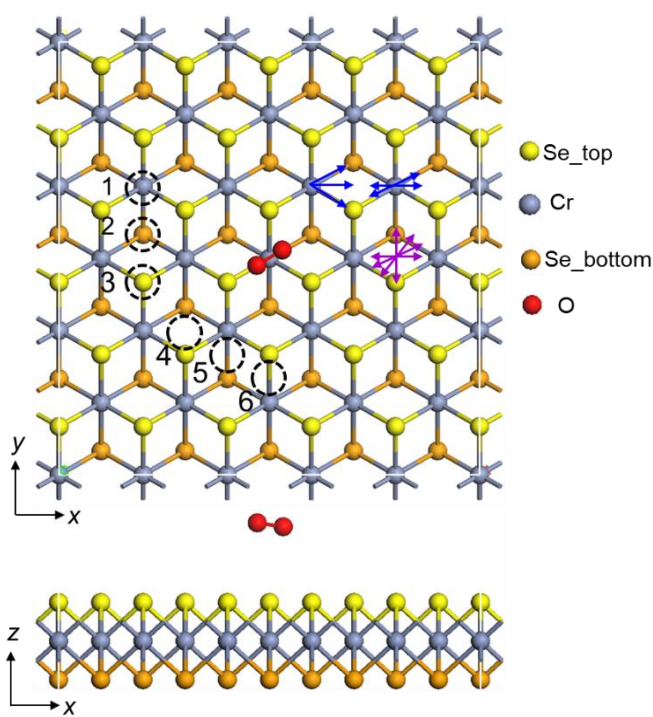

**Figure 1. Schematic model of the possible $O_2$ adsorption sites on monolayer $CrSe_2$ considered in our calculations.** Six adsorption sites for a $O_2$ molecule on $CrSe_2$ have been considered: (1) Cr top site, (2) bottom Se site, (3) top Se site, (4) Cr-bridge site, (5) top Se-bridge site, (6) bottom Se-bridge site. The top view and side view are shown in the upper and lower panels, respectively. The cyan, yellow, orange and red balls represent Cr, top Se, bottom Se and O atoms, respectively. The blue and purple arrow represent different orientation directions of $O_2$ on Cr top or Cr-bridge site.

Oxidation is one of most likely reasons for materials degradation in air[36, 37] and would significantly change atomic and electronic structures of 2D materials, thus impacting the key performance parameters, such as intrinsic magnetism and band structure[18]. We carried out DFT calculations to examine interacting details of $O_2$ molecules on monolayer $CrSe_2$. According to our experiences on black phosphorus degradation[38], the monolayer appears mostly easy for oxidation and we thus adopted a $CrSe_2$ monolayer for further calculations. A $5\times3\sqrt{3}$ rectangular supercell was adopted to model the surface upon $O_2$ adsorption, which ensures a separation of at least 15 Å between molecules. We carefully examined six possible adsorption sites (black dashed circles in Fig. 1) and different $O_2$ molecule orientations (colored arrows in Fig. 1), totally 27 kinds of $O_2$ adsorption configurations. On Cr and Se top site, we selected either one of O atoms or centroid of O₂ molecule as rotation center with a 15° step in x-y plane. On the bridge site, rotation angle of O$_2$ molecule is 0°, 30°, 60° and 90° in each configuration counterclockwise with respect to the x-axis. O$_2$ molecule cannot stably hold its initial position on bridge site during the relaxation process. As illustrated in Fig. 1, the most stable adsorption configuration of O$_2$ molecule is on Cr top site with an adsorption energy of -0.093 eV per molecule and a height of 3.14 Å. The angle between O$_2$ molecule orientation direction and x-axis is 30° in the most table configuration, which is at least 3 meV more stable than other configurations and is thus considered as initial state (IS) in the following discussions.

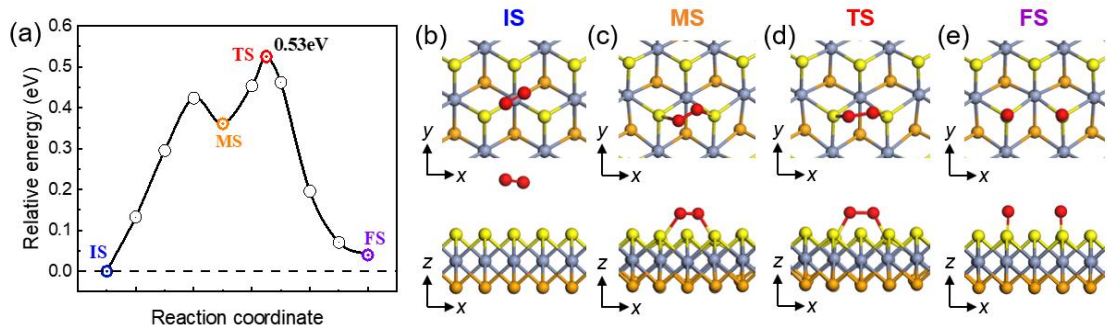

**Figure 2. Reaction pathway for O$_2$ molecule to dissociate two atoms on monolayer CrSe$_2$.** (a) Minimum energy path for O$_2$ dissociative adsorption on CrSe$_2$ obtained from CI-NEB calculations. Atomic configurations of (b) initial state (IS), (c) meta state (MS), (d) transition state (TS) and (e) final state (FS) configurations are presented in the schematic.

Figure 2 shows the initial state (IS), meta state (MS), transition state (TS), and final state (FS) of the reaction pathway for the O$_2$ dissociative adsorption revealed by the climbing-image nudged elastic band (CI-NEB) calculations. A horizontal movement of O$_2$ molecule to the top Se-bridge site plays a key role in transition from IS to MS, whose adsorption energy and height is 0.273 eV and 1.41 Å, respectively. Due to the bonding of Se atoms and O atoms in TS, the O-O bond is weakened with a length elongated to 1.48 Å compared to that of 1.23 Å of free O$_2$. Finally, the O-O bond breaks and two O atoms diffusing to two adjacent Se top sites to form final state. Energy barrier of the O$_2$ dissociation on pristine CrSe$_2$ is 0.53 eV (Fig. 2a). By comparing the relative energies of IS and FS, we found the dissociative adsorption of O$_2$ is endothermic with an energy gaining of 0.08 eV, which hinders the reaction occurring at normal conditions. These results show that the energetically unfavored surface oxidation reaction could be one of the reasons for the exceptional air-stability of monolayer CrSe$_2$ found in experiments[20].

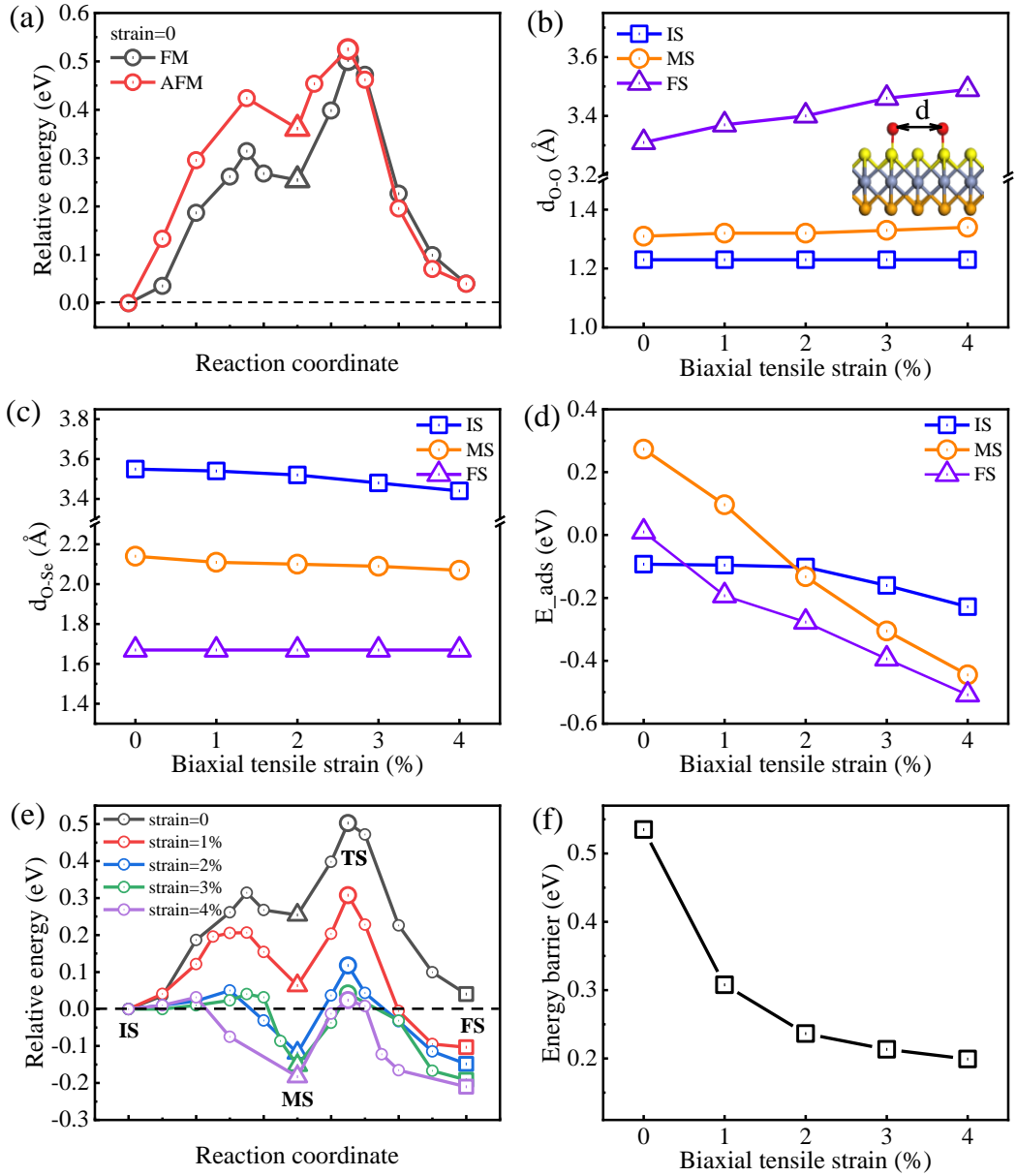

**Figure 3. Effect of biaxial tensile strain on atomic structure and surface oxidation reaction of CrSe$_2$ monolayer.** (a) Magnetic configuration effect on oxidation reaction in pristine CrSe$_2$. (b) O-O bond length, (c) distance between O atom and its nearest Se atom in IS, MS and FS as a function of strain. (d) Adsorption energies as a function of strain. (e) Reaction pathway for the dissociative adsorption of O$_2$ on monolayer CrSe$_2$ under biaxial tensile strain. MS, TS and FS are presented with triangles, circles and rectangles, respectively. (f) O$_2$ dissociation barriers as a function of biaxial strain.

The magnetic ground state of CrSe$_2$ may also affect its stability, so we calculated the oxidation process of CrSe$_2$ with ferromagnetic (FM) and antiferromagnetic (AFM) orders (Fig. 3a). The magnetic order of monolayer CrSe$_2$ has little influence on the surface oxidation reaction and is thus kept as FM in the following discussions. Epitaxy growth of 2D martials on substrate usually introduces interfacial strain because of lattice mismatch, which is long lasting and often plays a key role modifying the magnetic properties of 2D magnetism, such as exchange energy and magnetic anisotropy energy[16]. However, the role played by strain on air-stability of $CrSe_2$ is unclear yet. Therefore, we further discussed the influence of biaxial tensile strain. Under increasing biaxial tensile strain, the distance between O atoms in IS remains nearly constant (Fig. 3b). Due to the enhancement of O-Se hybridization in chemisorption (Fig. 3c), the O-O bond length increases from 1.31 Å to 1.34 Å in MS, which is conducive to the $O_2$ dissociation. As shown in Fig. 3d, the adsorption energy of MS and FS decrease faster than IS under increasing biaxial tensile strain. In-plane strain could stabilize the chemisorption configuration MS. The dissociative adsorption of $O_2$ becomes exothermic with an energy lowering of 0.1 eV under tensile strain reaching 1%. We carried out NEB calculations for the evolution of dissociated energy barriers under biaxial tensile strain (Fig. 3e). Apart from the endothermic to exothermic transition of $O_2$ dissociation, biaxial tensile strain in the range from 0 to 4% induces decrease of dissociative energy barrier from 0.53 to 0.20 eV. These results show that air-stability of monolayer $CrSe_2$ will decrease under biaxial tensile strain. Electronic structure can be used to understand the mechanism of strain tunability.

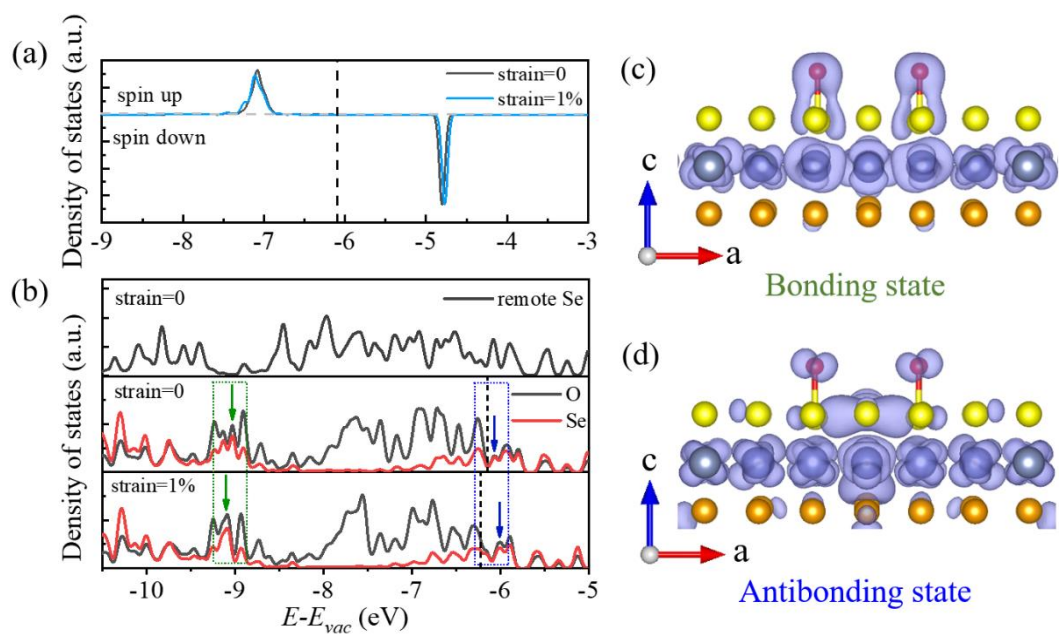

**Figure 4. Electronic structures of $O_2$/O atoms adsorbed on monolayer $CrSe_2$.** (a) Local density of states (LDOS) of $O_2$ molecule on $CrSe_2$ with and without applying tensile strain. The black

vertical dashed line is the Fermi level. (b) Projected density of states (PDOS) of O atom and bonding Se atom in FS under strain= 0 and 1%. The top panel shows the PODS of the Se atoms far away from the O atoms. The green and blue arrows are the bonding state and anti-bonding state of O-Se, respectively. The black vertical dashed line is the Fermi level. (c-d) Visualized wavefunction of O-Se bonding and antibonding states corresponding to the green and blue dashed box in Fig. 4b when strain = 1%, respectively. The isosurface value is set to 0.0004 $e$/Bohr$^3$.

In order to explain the transition from the endothermic to exothermic reaction, when biaxial tensile strain = 0 and 1%, we calculated the local density of states (LDOS) of molecule on CrSe$_2$ in IS and projected density of states (PDOS) of O and bonding Se in FS. From biaxial tensile strain = 0 to 1%, a variation of strain has little effect on the LDOS of O$_2$ molecule (Fig. 4a), which is consistent with the unchanged O-O bond length of O$_2$ physically adsorbed on CrSe$_2$ (Fig. 3b). As shown in Fig. 4b, energy splitting of O-Se bonding (green arrows) and antibonding (blue arrows) states under strain = 1% increase by about 0.2 eV compared with strain = 0. Energy splitting is also increased under increasing biaxial tensile strain. The result leads to a largely stabilized FS state and thus the endothermic to exothermic transition under strain over 1%.

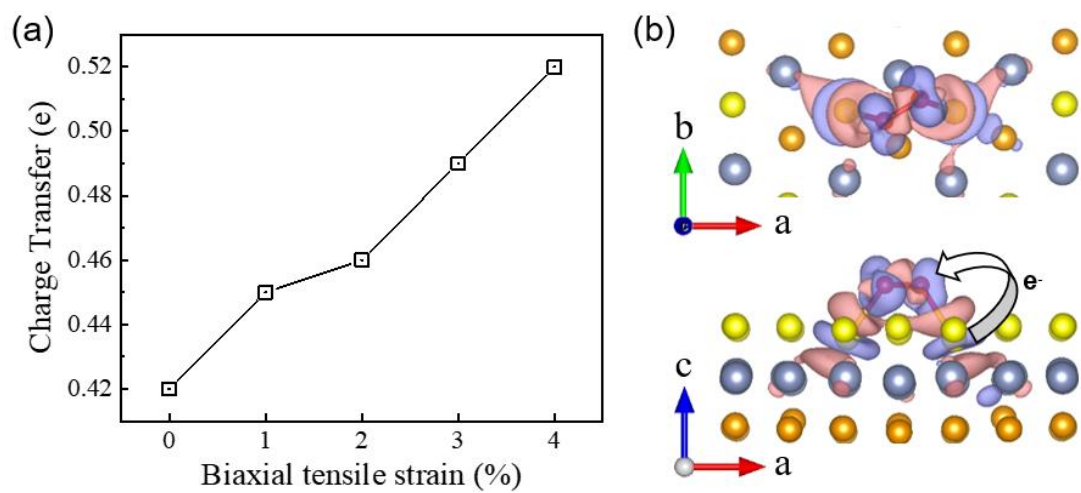

**Figure 5. Charge transfer from CrSe$_2$ to O$_2$.** (a) Effect of biaxial tensile strain on charge transfer from monolayer CrSe$_2$ to O$_2$. (b) Differential charge density of O$_2$@CrSe$_2$ at strain = 1%. The purple and light red contours indicate charge accumulation and charge reduction after O$_2$ adsorption, respectively. The isosurface value is set to 0.003 $e$/Bohr$^3$.

Previous studies have shown that charge transfer between adsorbed molecule and surface of materials can help to understand the mechanism of surface oxidation reaction[39, 40]. In the surface oxidation process of $O_2$ on $CrSe_2$, the MS is crucial to $O_2$ dissociation. Molecules adsorbed to the surface get electrons filled in antibonding state, which contribute to molecule dissociation[41]. Bader charge analysis was used to obtain the number of transferred electrons from $CrSe_2$ to $O_2$ in MS. Schematic diagram of charge transfer from $CrSe_2$ to $O_2$ is shown in (Fig. 5b). Applied biaxial tensile strain in the range from 0 to 4% induced a decrease of work function of monolayer $CrSe_2$ from 6.44 to 6.18 eV. Due to a decrease of work function, the electrons transferred from $CrSe_2$ to $O_2$ induced an increase from 0.42 to 0.52 e in process of MS to TS. The increase of transferred electrons could activate the O-O bond and facilitate its rupture, leading the reduction of energy barriers.

## Conclusion

In conclusion, we investigated the dissociation process of $O_2$ on $CrSe_2$ to understand the air-stability mechanism and strain tunability. In this work, our first-principles calculations show that the process of $O_2$ dissociating into O atoms on pristine $CrSe_2$ is an endothermic reaction. The energetically unfavored surface oxidation reaction could be one of the reasons for the exceptional air-stability of $CrSe_2$ measured in experiment. We also found biaxial tensile strain has a negative effect on air-stability of monolayer $CrSe_2$. The dissociation reaction of $O_2$ becomes an exothermic reaction owing to the nearly unchanged IS and stabilized FS state under biaxial tensile strain. The energy barrier of $O_2$ is reduced under strain, which could be attributed to enhanced charge transfer from $CrSe_2$ to $O_2$. In the preparation and devices application of 2D $CrSe_2$, the stretch of $CrSe_2$ should be avoided to maintain its air-stability.

## Acknowledgement

We gratefully acknowledge financial support from the Ministry of Science and Technology (MOST) of China (Grant No. 2018YFE0202700), the National Natural Science Foundation of China (Grants No. 61761166009, No. 11974422 and No. 12104504), and the Strategic Priority Research Program of Chinese Academy of Sciences (Grant No. XDB30000000). C.W. was supported by the China Postdoctoral Science Foundation (2021M693479). Calculations were performed at the Physics Lab of High-Performance Computing of Renmin University of China, Shanghai Supercomputer Center.

## References

1. Jimenez, V. O. et al. A magnetic sensor using a 2D van der Waals ferromagnetic material. *Sci. Rep.-UK*. **10** (2020).
2. Chappert, C., Fert, A. & Van Dau, F. N. The emergence of spin electronics in data storage. *Nat. Mater.* **6**: 813-823 (2007).
3. Burch, K. S., Mandrus, D. & Park, J. Magnetism in two-dimensional van der Waals materials. Nature. **563**: 47-52 (2018).
4. Gong, C. & Zhang, X. Two-dimensional magnetic crystals and emergent heterostructure devices. *Science*. **363**: 706 (2019).
5. Gibertini, M., Koperski, M., Morpurgo, A. F. & Novoselov, K. S. Magnetic 2D materials and heterostructures. Nat. Nanotechnol. **14**: 408-419 (2019).
6. Lin, X., Yang, W., Wang, K. L. & Zhao, W. Two-dimensional spintronics for low-power electronics. *Nat. electron*. **2**: 274-283 (2019).
7. Guo, Y. et al. Magnetic two-dimensional layered crystals meet with ferromagnetic semiconductors. *InfoMat.* **2**: 639-655 (2020).
8. Huang, B. et al. Layer-dependent ferromagnetism in a van der Waals crystal down to the monolayer limit. Nature. **546**: 270-273 (2017).
9. Zheng, F. et al. Tunable spin states in the two-dimensional magnet $CrI_3$. *Nanoscale*. **10**: 14298-14303 (2018).
10. Webster, L. & Yan, J. Strain-tunable magnetic anisotropy in monolayer $CrCl_3$, $CrBr_3$ and $CrI_3$. *Phys. Rev. B*. **98**: 144411 (2018).
11. Gong, C. et al. Discovery of intrinsic ferromagnetism in two-dimensional van der Waals crystals. Nature. **546**: 265-269 (2017).
12. Ostwal, V., Shen, T. & Appenzeller, J. Efficient spin-orbit Torque switching of the semiconducting van der Waals ferromagnet $Cr_2Ge_2Te_6$. Adv. Mater. **32**: 1906021 (2020).
13. Deng, Y. et al. Gate-tunable room-temperature ferromagnetism in two-dimensional $Fe_3GeTe_2$. Nature. **563**: 94-99 (2018).
14. Fei, Z. et al. Two-dimensional itinerant ferromagnetism in atomically thin $Fe_3GeTe_2$. *Nat. Mater.* **17**: 778-782 (2018).
15. Cong Wang et al. Layer and doping tunable ferromagnetic order in two-dimensional CrS2 layers. *Phys. Rev. B*. **97**: 245409 (2018).
16. Linlu, W. et al. In-plane epitaxy strain tuning intralayer and interlayer magnetic couplings in CrSe2 and CrTe2 mono- and bi-layers. *arXiv:2202.11956* (2022).
17. Sivadas, N. et al. Stacking-Dependent Magnetism in Bilayer CrI3. Nano Lett. **18**: 7658-7664 (2018).
18. Zhang, T. et al. Degradation chemistry and kinetic stabilization of magnetic $CrI_3$. J. Am. Chem. Soc. (2022).
19. Su, C. et al. Waterproof molecular monolayers stabilize 2D materials. *Proceedings of the National Academy of Sciences*. **116**: 20844-20849 (2019).
20. Li, B. et al. Van der Waals epitaxial growth of air-stable $CrSe_2$ nanosheets with thickness-tunable magnetic order. Nat. Mater. **20**: 818-825 (2021).
21. Ramos, M. et al. Ultra-broad spectral photo-response in $FePS_3$ air-stable devices. *NPJ 2D Mater. Appl.* **5**: 1-9 (2021).
22. Telford, E. J. et al. Layered antiferromagnetism induces large negative magnetoresistance in the van der Waals semiconductor CrSBr. Adv. Mater. **32**: 2003240 (2020).

23. Meng, L. et al. Anomalous thickness dependence of Curie temperature in air-stable two-dimensional ferromagnetic 1T-CrTe2 grown by chemical vapor deposition. Nat. Commun. **12**: 809 (2021).
24. Perdew, J., Burke, K. & Ernzerhof, M. Generalized gradient approximation made simple. *Phys. Rev. Lett.* **77**: 3865-3868.
25. Blöchl, P. E. Projector augmented-wave method. *Phys. Rev. B*. **50**: 17953 (1994).
26. G. Kresse & Joubert, D. From ultrasoft pseudopotentials to the projector augmented-wave method. *Phys. Rev. B*. **59**: 1758 (1999).
27. Kresse, G. & Furthmuller, J. Efficient iterative schemes for ab initio total-energy calculations using a plane-wave basis set. *Phys. Rev. B*. **54**: 11169-11186 (1996).
28. Grimme, S., Antony, J., Ehrlich, S. & Krieg, H. A consistent and accurate ab initio parametrization of density functional dispersion correction (DFT-D) for the 94 elements H-Pu. *J. Chem. Phys.* **132**: 154104 (2010).
29. Cococcioni, M. & de Gironcoli, S. Linear response approach to the calculation of the effective interaction parameters in the LDA+U method. *Phys. Rev. B*. **71**: 035105 (2005).
30. Wang, C. et al. Bethe-Slater-curve-like behavior and interlayer spin-exchange coupling mechanisms in two-dimensional magnetic bilayers. *Physical review. B*. **102** (2020).
31. Heyd, J., Scuseria, G. E. & Ernzerhof, M. Hybrid functionals based on a screened Coulomb potential. *The Journal of Chemical Physics*. **118**: 8207-8215 (2003).
32. Jiang, P. et al. Stacking tunable interlayer magnetism in bilayer CrI3. *Physical review. B*. **99**: 144401 (2019).
33. Mills, G., Jónsson, H. & Schenter, G. K. Reversible work transition state theory: application to dissociative adsorption of hydrogen. *Surf. Sci.* **324**: 305-337 (1995).
34. Olsen, R. A. et al. Comparison of methods for finding saddle points without knowledge of the final states. J. Chem. Phys. **121**: 9776-9792 (2004).
35. Tang, W., Sanville, E. & Henkelman, G. A grid-based Bader analysis algorithm without lattice bias. *J. Phys.: Condens. Matter*. **21**: 084204 (2009).
36. Zhao, Y. et al. High-electron-mobility and air-stable 2D layered $PtSe_2$ FETs. Adv. Mater. **29**: 1604230 (2017).
37. Guo, Y., Zhou, S., Bai, Y. & Zhao, J. Oxidation Resistance of Monolayer Group-IV Monochalcogenides. *ACS Appl. Mater. Inter.* **9**: 12013-12020 (2017).
38. Yuan Huang, J. Q. K. H. Interaction of Black Phosphorus with Oxygen and Water. *Chem. Mater.* **28**: 8330-8339 (2016).
39. Kistanov, A. A. et al. A first-principles study on the adsorption of small molecules on antimonene: oxidation tendency and stability. *Journal of materials chemistry. C, Materials for optical and electronic devices*. **6**: 438-4317 (2018).
40. Zhou, C. et al. Effect of external strain on the charge transfer: Adsorption of gas molecules on monolayer GaSe. Mater. Chem. Phys. **198**: 49-56 (2017).
41. Rao, Y. C. & Duan, X. M. Pd/Pt embedded CN monolayers as efficient catalysts for CO oxidation. *Phys. Chem. Chem. Phys.* **21**: 25743-25748 (2019).